\documentstyle[11pt]{article}

\makeatletter
\def\dif{{\rm d}}
\def\deriv{\@ifnextchar[{\@deriv}{\@deriv[]}}
   \def\@deriv[#1]#2#3{\mathchoice%
{{\dif^{#1}#2\over\dif{#3}^{#1}}}{{\dif^{#1}#2/\dif{#3}^{#1}}}%
{{\dif^{#1}#2\over\dif{#3}^{#1}}}{{\dif^{#1}#2/\dif{#3}^{#1}}}}
\def\derpar#1#2{\mathchoice%
{{\partial#1\over\partial#2}}{{\partial#1/\partial#2}}%
{{\partial#1\over\partial#2}}{{\partial#1/\partial#2}}}

\def\secteqno{\@addtoreset{equation}{section}%
\def\theequation{\thesection.\arabic{equation}}}
\newcounter{subequation}
\def\thesubequation{\alph{subequation}}
\def\sneqnarray{\stepcounter{equation}\let\@currentlabel=\theequation
\setcounter{subequation}{1}
\def\@eqnnum{{\rm (\theequation.\thesubequation)}}
\global\@eqcnt\z@\tabskip\@centering\let\\=\@eqncr\let\@@eqncr=\@@sne
qncr
$$\halign to \displaywidth\bgroup\@eqnsel\hskip\@centering
 $\displaystyle\tabskip\z@{##}$&\global\@eqcnt\@ne
 \hskip 2\arraycolsep \hfil${##}$\hfil
 &\global\@eqcnt\tw@ \hskip 2\arraycolsep $\displaystyle\tabskip\z@{##}$\hfil
  \tabskip\@centering&\llap{##}\tabskip\z@\cr}
\def\endsneqnarray{\@@sneqncr\egroup $$\global\@ignoretrue}
\def\@@sneqncr{\let\@tempa\relax
   \ifcase\@eqcnt \def\@tempa{& & &}\or \def\@tempa{& &}
   \else \def\@tempa{&}\fi
     \@tempa \if@eqnsw\@eqnnum\stepcounter{subequation}\fi
     \global\@eqnswtrue\global\@eqcnt\z@\cr}
\def\nobiblabels{\def\@lbibitem[##1]##2{\@bibitem{##2}}}
\makeatother
\def\tabaddress#1{{\small\it\begin{tabular}[t]{c}#1\\[1.2ex]\end{tabular}}}

\def\ben{\begin{enumerate}}
\def\een{\end{enumerate}}
\def\beq{\begin{equation}}
\def\eeq{\end{equation}}
\def\bea{\begin{eqnarray}}
\def\eea{\end{eqnarray}}
\def\beann{\begin{eqnarray*}}
\def\eeann{\end{eqnarray*}}
\def\beasn{\begin{sneqnarray}}
\def\eeasn{\end{sneqnarray}}

\def\UBECM{Departament d'Estructura i Constituents de la Mat\`eria\\
   Universitat de Barcelona\\
   Av.~Diagonal 647\\
   08028 Barcelona\\
   Catalonia, Spain}

\begin{document}
\thispagestyle{empty}
\title{Equivalence of Faddeev-Jackiw
and Dirac approaches for gauge theories}
\author{\sc
J. Antonio Garc\'ia\thanks{On leave from Instituto de Ciencias Nucleares,
Universidad Nacional Aut\'onoma de M\'exico,
Apartado Postal 70-543,
04510 M\'exico, D.F.  E-mail: antonio@teorica0.ifisicacu.unam.mx}\,
and
Josep M. Pons
\\
\def\baselinestretch{1.2}
\tabaddress{\UBECM}
}
\date{}

\def\ppclassif{{\noindent\tt UB-ECM-PF 95/15\\hep-th/9610067}}

\pagestyle{myheadings}
\markright{\sc J.A. Garc\'ia, J.M. Pons,
\rm `Equivalence of  F-J
and Dirac...'}

\def\baselinestretch{1.2}
\nobiblabels
\secteqno

\maketitle

\begin{abstract}

The equivalence between the Dirac method and Faddeev-Jackiw analysis for
gauge theories is proved.  In particular we trace out, in a stage by
stage procedure, the standard classification of first and second class
constraints of Dirac's method in the F-J approach.  We also find that
the Darboux transformation implied in the F-J reduction process can be
viewed as a canonical transformation in Dirac approach.  Unlike Dirac's
method the F-J analysis is a classical reduction procedure, then the
quantization can be achieved only in the framework of reduce and then
quantize approach with all the know problems that this type of
procedures presents.  Finally we illustrate the equivalence by means of
a particular example.
\end{abstract}

\vfill
\ppclassif
\clearpage

%%%%%%%%%%%%%%%%%%%%%%%%%%%%%%%%%%%%%%%%
%%%%%%%%%%%%%%%%%%%%%%%%%
\section{Introduction}

The classical treatment of the dynamics of gauge theories (including
those with repara\-metrization invariance) described by a variational
principle was first solved by Dirac \cite{Dirac} and Bergmann
\cite{Bergmann} in the early 50's. Dirac was mainly interested in
the Hamiltonian approach to general relativity, and his leading efforts
in this field were completed in the 60's with the ADM formalism
\cite{ADM}.  The application of Dirac's work to gauge theories like EM
or YM came later in the 70's (for instance, the basic canonical
commutations for the electromagnetic potential in Coulomb gauge, first
obtained through heuristic methods, were then understood \cite{HTR} as a
direct application of the Dirac bracket).  Also in the 70's the
geometrization of Dirac's method was successfully addressed \cite{Gotay}
and in the 80's there appeared general proofs of the equivalence of the
(classical) Hamiltonian and Lagrangian treatment for gauge theories
\cite{Barcelona}.  With regard to the quantization program, Dirac's
approach opened the way to the so called Dirac method of quantization,
where constraints are implemented as operators in Hilbert space.  The
equivalence of this method with the so called {\it reduced} quantization
(i.e., when the classical degrees of freedom are eliminated before
quantization) is still a subject of controversy \cite{Reduced}.  Let us
mention also the discovery in the 70's of the BRST symmetry \cite{BRST},
which appeared as a reminiscence of the classical gauge symmetry after
fixing the gauge through Faddeev-Popov procedure \cite{FP}.  In this
sense, the classical approach by Dirac had its quantum continuation
through the BRST methods.  These methods have provided the most powerful
tool, the field-antifield formalism, to quantize any kind (reducible
gauge algebra, soft algebra, open algebra, etc.) of local gauge field
theory \cite{FA}.

The main features of Dirac approach, either in Lagrangian or Hamiltonian
formalism are a) the possibility to keep all the variables in phase
space or velocity space, b) the construction of an algorithm to
determine, through a step by step procedure, the final constraint
surface where the motion takes place, c) the elucidation of the true
degrees of freedom of the theory, separated from the gauge --and hence
unphysical-- ones.  In general, quotienting out the gauge degrees of
freedom can be done by ``fixing the gauge'', which amounts to the
introduction of a new set of constraints from the outset, the Gauge
Fixing constraints.

More recently, a new method to classically deal with gauge theories was
devised through the work by Faddeev and Jackiw\cite{FJ}.  Unlike
Dirac's, in the F-J method, the variables are reduced to the physical
ones and, at least formally, the procedure looks simpler.  We have
considered that it is worth to explore F-J method to see to what extent
it differs --if it does-- from Dirac's, and what are the advantages of
each one.  Let us immediately state our conclusion: though technically
different, both methods are equivalent, as it should be.  The main
difference in procedure is that F-J method is a method of {\it
reduction} to the physical degrees of freedom --and its variational
principle.

It is true nevertheless that Dirac's method also includes the
possibility to eliminate variables.  In fact, the adoption of the
Dirac's bracket (which is the bracket associated to the symplectic form
that one can define in the second class constraint surface as the
projection of the symplectic form in phase space) allows for the
elimination of a variable for every second class constraint.  Instead,
in the F-J procedure, even in the case when some first class constraints
are present, the reduction can be still performed, although the
projected symplectic form becomes degenerate.  This implies that some
dynamical variables and its equations of motion may be lost.  A key
feature of F-J procedure is that this fact does not affect the physical
content of the theory.

As we will see in the next section, F-J procedure focuses exclusively 
getting the equations of motion (or, equivalently, its
associated variational principle) for the set of physical variables,
discarding anything else.  Some of the variables which in Dirac's
approach are related to the physical ones through constraints are
quickly eliminated in F-J method together with other variables which are
going to become gauge degrees of freedom.  This fact  explains
the efficiency of F-J method: It does not produces superfluous
information that is going to be discarded later on, contrary to what
happens in Dirac's, where we can keep this superfluous information till
the end.

Obviously, this efficiency must pay some price. Apart from the
technically difficult Darboux transformation implied in the
F-J reduction process, the indiscriminate
elimination of variables leads to the typical difficulties that plague
the reduction procedures: the general lost of covariance and in some
cases even the lost of properties of locality for the reduced field
theory.  Let us mention that the BRST quantization methods mentioned
above lie just on the ``other side'' of the ``reductionist'' approach: In
any version of the BRST formalism, the set of variables (the original
fields in the theory), instead of being reduced (thus loosing
covariance, or locality) is enlarged with new variables (ghosts,
antifields,..).

Beyond these issues of covariance or locality there is the important
phenomena of quantum gauge anomalies, i.e., classical gauge symmetries
that, due to the regularization procedures involved in the quantization
program, do not become quantum symmetries.  In such a case it is clear
that one must be very careful in the classical reduction, since it
cannot take quantum gauge anomalies into account, so they must show
up somewhere else (for instance, in the early methods for the reduced
quantization of the bosonic string, the disappearance of the Weyl
anomaly was the hidden cause that the Poincare group was only correctly
realized at 26 spacetime dimensions).

A covariant analysis of the reduced phase space (covariant symplectic
approach) has been worked in \cite{Witten}.  For a comparative
examination of this approach with the Hamiltonian reduction and the
Peierls bracket procedure see \cite{H1}.  The relation between first
order Lagrangians and Dirac brackets in Darboux coordinates was
worked out in \cite{sh-lfu}. On the other hand the results obtained
from F-J approach has been compared with the corresponding results of
Dirac method for the case of constant symplectic matrix through simple
particular examples in \cite{KM} without proving the equivalence
between the two approaches.  An analysis from the point of view of the
so called symplectic reduction procedure has been performed by the
authors of \cite{BW}, avoiding the difficult Darboux transformation
implied in the original analysis of F-J by expanding, at each stage of
the algorithm, the number of variables in the phase space.  Some
applications of F-J approach can be found in \cite{{A1},{A2},{A3},{A4}}
and the extension to the fermionic case for constant symplectic
structures is accomplished in \cite{G}.  Recently the authors \cite{MZ}
compared the Dirac quantization program of promoting the Dirac brackets
to commutators with the resulting brackets in the F-J approach using a
gauge fixing procedure.

The purpose of this paper is to analyze and explain in detail the
equivalence of Dirac's method with Faddeev and Jackiw's one.  It is usually
said that with F-J method, the separation of constraints in first and
second class, customary in Dirac's method, disappears.  We will see that
this is not completely true and that it is possible in the F-J method
to keep track, at least partially, of the standard Dirac's
classification of constraints as first class and second class.

The difference between the two stabilization procedures, that of Dirac
and F-J, is that in the first case we obtain a new symplectic form (the
Dirac brackets) and some first class constraints at each stage of the
reduction procedure.  The stabilization algorithm terminates when no new
constraints appear in the theory.  We end up with a gauge theory with
only first class constraints in a partially reduced phase space.
The final reduction is now performed by means of the gauge fixing
procedure.  In the F-J case, at each stage of the algorithm, we plug the
constraints into the action and diagonalize (by means of a Darboux
transformation) the symplectic form obtaining as a byproduct new
constraints that by means of the reduction process result in a new
symplectic structure that may be degenerate.  The procedure continues
through a new diagonalization and ends up with a non degenerate
symplectic structure that actually represents the Dirac brackets in the
reduced phase space.

To compare the two procedures we will work in coordinates that allow for
a canonical representation of the constraint surface at each stage of
the stabilization procedure.  In these coordinates the Dirac brackets
become diagonal at each stage of the algorithm.  We assume that all the
constraints that appear in the formalism are effective, i.e., its
gradient on the constraint surface is different from zero. In fact, this
assumption is crucial and its failure may obstruct the equivalence
between Dirac's and Faddeev and Jackiw's methods.

The equivalence of Dirac's with Faddeev and Jackiw's method is proved in
section 2. An example are given is section 3 and section 4 is devoted to
conclusions.

%%%%%%%%%%%%%%%%%%%%%%%%%%%%%%%%%%%%%%%%
%%%%%%%%%%%%%%%%%

\section{The equivalence of Dirac's and F-J methods}

We will take as starting point a canonical Hamiltonian $H_c(q,p)$ and a
set of primary constraints $\phi_\mu = 0$.  One can consider the phase
space as the new configuration space for the canonical Lagrangian,
\beq
L = p_i \dot q^i - H_c(q,p) - \lambda^\mu \phi_\mu.
\label{L}
\eeq

$\lambda^\mu$ are Lagrange multipliers which are taken as arbitrary
functions of time --until some of them get determined by consistency
requirements.  The Euler-Lagrange equations for $L$ yield the usual
Hamilton-Dirac equations on the surface of primary constraints $\phi_\mu
=0$.

\subsection{F-J analysis}

In F-J method one eliminates as many variables as constraints $\phi_\mu$
just by plugging these constraints --which now are holonomic-- into the
Lagrangian.  If we call $x^s$ the remaining set of variables, the
reduced Lagrangian will take the form\footnote{Notice that the sub or
super-index carries information both on the type of variable --labeled
by the letter-- under consideration and the range of values that it can
take. }
\beq
L' = a_s(x)\dot x^s - H(x),
\label{La}
\eeq
for some functions $a_s$ and $H$.
At this point we can perform a Darboux transformation,\footnote{For a careful
analysis of the Darboux theorem
for singular systems, see \cite{Olver}.}
\beq
x^s \longrightarrow Q^r,\, P_r ,\, Z_a
\label{darboux}
\eeq
such that $L'$ takes again the form of a canonical Lagrangian:
\beq
L' = P_r \dot Q^r - H'(Q^r,P_r,Z_a)
\label{L'}
\eeq

Notice that in general there appear a certain number of variables,
$Z_a$, without its canonical counterpart.  This reflects the fact that
the primary constraint surface can be endowed with a pre-symplectic
structure, and not a symplectic one.  These variables $Z_a$ play the
role of auxiliary variables.  By using its equations of motion, some of
them can be eliminated.  Let us name $Z_{a_1}$ the maximum set of
variables that can be eliminated, and let $Z_{a_2}$ be the rest.  In
other words, the equations of motion for the variables $Z_a$,
\beq
\derpar{H'}{Z_a} = 0,
\label{H'}
\eeq
allow for the isolation of $Z_{a_1}$ in terms of the rest of variables,
\beq
Z_{a_1} = f_{a_1} (Q^r,P_r,Z_{a_2}),
\label{zeta1}
\eeq
together with some other possible relations which are free from these
$Z_a$ variables,
\beq
f_{a_2} (Q^r,P_r) = 0.
\label{zeta2}
\eeq
(some of these last relations --or maybe all-- may be empty).  Here
$a_2$ denotes the maximum number of independent relations of type
$(\ref{zeta2})$.

Now let us proceed to a new reduction: the elimination of the variables
$Z_{a_1}$ by plugging (\ref{zeta1}) into (\ref{L'}). In this way we get
\beq
L_R = P_r \dot Q^r - H''(Q^r,P_r,Z_{a_2}).
\eeq

But now the dependence of $H''$ with respect to $Z_{a_2}$ cannot be more
than linear --just by construction-- because otherwise variables of the
type $Z_{a_2}$ could still be eliminated through the procedure above, in
contradiction with the fact that we had already chosen the maximum set
of variables that could be eliminated, therefore
\beq
L_R = P_r \dot Q^r - H_R(Q^r,P_r) - Z_{a_2} f_{a_2} (Q^r,P_r).
\label{LR}
\eeq

Note that $f_{a_2}$ is here an equivalent representation of the
constraints defined by $(\ref{zeta2})$ and that we have arrived at a
reduced Lagrangian, in a reduced phase space, that have the same form as
$(\ref{L})$.  This finishes one step --the first-- in F-J algorithm.
The variables $Z_{a_2}$ play from now on the role of Lagrange
multipliers.  Now the procedure is repeated again and again until the
disappearance of these $Z$-type variables from the formalism.  The
remaining variables will be the physical ones, associated to the true
degrees of freedom of the system.  Equivalently the procedure ends when
we obtain a non degenerate symplectic matrix from the reduced
Lagrangian.  This symplectic matrix represents the Dirac brackets in the
reduced phase space.

Now we are ready to start the Dirac's procedure, adapted for a
transparent comparison with the F-J approach.  So let us go back to
(\ref{L}).  The constraints $\phi_\mu = 0$ can be classified into first
and second class.  Then it is possible to change to an equivalent set of
constraints (i.e., a change of the basis of functions that generate the
ideal of functions vanishing at the primary constraint
surface),\footnote{It is worth noting that, in the case of first class
constraints, the BRST formalism in the extended phase space (including
ghost, antighosts ...), the transition from one set of constraints to an
equivalent one $(\ref{M})$ is induced by a canonical transformation in
the extended phase space.  This fact is at the root of the well known
result that states that the BRST charge is unique up to a canonical
transformation in the extended phase space. \cite{HT}.}
\beq
\phi_\mu \rightarrow \xi_\mu = M_\mu^\nu (q,p) \phi_\nu,\qquad \det M
\neq 0,
\label{M}
\eeq
with $\xi_\mu = \varphi_a,\, \chi_l,\, \bar{\chi}_l$, and such that
\bea
&&\{\varphi_a,\,\chi_{a'} \} = \{\varphi_a,\,\chi_l \} =
\{\chi_l,\,\chi_{l'}\} = \\
&&\{\varphi_a,\,\bar{\chi}_l \} =
\{\bar{\chi}_l,\,\bar{\chi}_{l'} \} = 0, \\
&&\{\chi_l,\,\bar{\chi}_{l'} \} = \delta_{l,l'}.
\eea

Notice that the assumption that all constraints are effective
is crucial for this change of basis to exist.
A canonical transformation can bring these constraints to canonical
coordinates,
\beq
(q_i,\,p^i) \rightarrow (Q^i,\,P_i) = (Q^r,\,P_r,\, Q^a,\,P_a,\,
Q^l,\,P_l,),
\label{cr}
\eeq
with
$$
P_a = \varphi_a, \quad Q^l = \chi_l, \quad  P_l = \bar{\chi}_l.
$$

We can also redefine the Lagrange multipliers $\eta^\mu =
{M^{-1}}_\nu^\mu
\lambda^\nu$.  Then the original Lagrangian can be written with the new
variables,
\beq
L = P_i \dot Q^i - \bar{H}_c(Q^i,P_i)
- \eta^a P_a -\eta^l P_l - \bar{\eta}^l Q^l,
\label{nou-L}
\eeq
where we have discarded a total derivative $\dot{F}(Q,P)$ coming from
$p_i \dot q^i = P_i \dot Q^i + \dot{F}(Q,P)$.  The action induced in the
reduced phase space differ weakly from the action in all phase space by
the boundary term $F(Q,P)$ evaluated at the extremal points that are
fixed in the variational principle.  It is important to note that this
term may not be invariant under the gauge transformations generated by
the first class constraints\cite{HT}.  For the problems that can appear
with the consistence between the boundary terms and the gauge fixing, at
the level of the variational principle, see\cite{HTV}.

With this ``canonical representation of the primary constraint
surface'', the F-J method can be constructed as follows.  Plugging the
constraints $P_a = 0,\,Q^l = 0,\, P_l = 0$, into (\ref{nou-L}) we get
\beq
L' = P_r \dot Q^r - H'(Q^r,P_r,Q^a)
\label{LRFJ}
\eeq
with $H'(Q^r,P_r,Q^a) = \bar{H}_c(Q^r,P_r,Q^a,P_a=0,Q^l=0,P_l=0)$.  So
we have arrived directly to (\ref{L'}) and the method continues as
before.  Notice that we have identified the variables $Z_a$ as $Q^a$,
the conjugate variables to the primary first class constraints $P_a$.

Note that the Lagrangians $(\ref{LRFJ})$ and $(\ref{L'})$ coincide.
This fact can be explained as follows.  The Darboux transformation
$(\ref{darboux})$, essentially no-canonical, shows up as a canonical
transfromation in a bigger space --the phase space in which the
constraint surface is embedded--.  This is a particular instance of how
the addition of new variables can give us more symmetry.  In this case a
non-canonical transformation is converted into a canonical one.
Extending off the constraint surface the Darboux transformation we get a
canonical transformation.  Obviously the extension must be zero {\it on}
the constraint surface and this in turn imply that the extended
variables must be a combination of the constraint functions as in
$(\ref{cr})$.

%%%%%%%%%%%%%%%%%%%%%%%%%%%%%%%

\subsection{Dirac analysis}

Consider now the Dirac's method. Having used the change of basis of
constraints and the canonical transformation above, we can start with
the canonical Hamiltonian  $\bar{H}_c(Q^i,P_i)$ and the constraints
$P_a = 0,\,Q^l = 0,\, P_l = 0$. The initial Dirac Hamiltonian is
\beq
\bar{H}_c(Q^i,P_i) + \eta^a P_a +\eta^l P_l +\bar{\eta}^l Q^l.
\eeq

Applying the stabilization algorithm to the second class constraints
$Q^l, \,P_l$ we determine $\eta^l = 0,\, \bar{\eta}^l = 0$.  Now a
trivial implementation of the Dirac bracket allows for the elimination
of the second class constraints.  The new Dirac Hamiltonian, in  the
partially reduced phase space --for the variables $Q^r,\,P_r,\,Q^a,\,P_a$--, is
\beq
H'(Q^r,P_r,Q_a) + \eta^a P_a,
\eeq
with $H'(Q^r,P_r,Q_a) = \bar{H}_c(Q^r,P_r,Q^a,P_a=0,Q^l=0,P_l=0)$. We
are entitled to put $P_a=0$ within the canonical Hamiltonian because
this Hamiltonian is only unambiguously defined on the surface of
primary constraints. Now consider the stabilization algorithm for the
primary first class constraints $P_a$. We get the secondary constraints
\beq
\dot{P}_a = \{ P_a,\, H' \} = - \derpar{H'}{Q^a} = 0,
\label{sec}
\eeq
which are nothing but (\ref{H'}).  We can split, as we did before, these
secondary constraints between those that allow for the elimination of
the variables $Q^a$ and the rest.  Using again the splitting $a =
(a_1,\,a_2)$, and having (\ref{H'}) and
(\ref{zeta1}) in mind, (\ref{sec}) can be written as
\beq
Q^{a_1} - f_{a_1} (Q^r,P_r,Q^{a_2}) = 0, \qquad
f_{a_2} (Q^r,P_r) = 0.
\label{split}
\eeq

Now let us partially proceed to the second step in Dirac's method.  It
is well known that each step comprises the possible determination of
some Lagrange multipliers and the possible introduction of new
constraints.  In our case it is immediate to see that the stabilization
of the first set of secondary constraints in (\ref{split}), $Q^{a_1} -
f_{a_1} (Q^r,P_r,Q^{a_2}) = 0$, leads to the determination of the
Lagrange multipliers $\eta_{a_1}$, whereas the stabilization of the
second set, $f_{a_2} (Q^r,P_r) = 0$, leads to the potential appearance
of new --tertiary-- constraints.  In Dirac's language, the constraints
$Q^{a_1} - f_{a_1} (Q^r,P_r,Q^{a_2}) = 0$ induce the change of status of
some part of the up-to-now primary first class constraints, $P_a=0$,
which will become second class.  More specifically, $P_{a_1} =0$ are the
constraints that become second class.  With this new set of second class
constraints, $P_{a_1} =0,\, Q^{a_1} - f_{a_1} (Q^r,P_r,Q^{a_2}) = 0$, we
can take again the Dirac bracket and get rid of the canonical variables
$P_{a_1} ,\, Q^{a_1}$.  Thus we are led to the newly reduced Dirac
Hamiltonian
\beq
H''(Q^r,P_r,Q^{a_2}) + \eta^{a_2} P_{a_2},
\eeq
in a phase space with canonical variables $Q^r,\,P_r,\,Q^{a_2},\,
P_{a_2}$. $H''$ is $H'$ with the substitution of $Q^{a_1}$ using
(\ref{split}).
The argument developed before tells us that the dependence of $H''$
with respect to $Q^{a_2}$ is at most linear. The Dirac Hamiltonian is,
therefore,
\beq
H_R(Q^r,P_r) + Q^{a_2} f_{a_2} (Q^r,P_r) + \eta^{a_2} P_{a_2},
\label{HRD}
\eeq

Notice that the last summand only have effect in the equations of motion
for the variables $Q^{a_2}$, which become, as of now, arbitrary
functions:
\beq
\dot{Q}^{a_2} = \eta^{a_2},
\label{primit}
\eeq

This special fact allows for a further reduction that it is not usually
performed in the Dirac's method: we can eliminate the last summand in
$(\ref{HRD})$,
reinterpret $Q^{a_2}$ as new Lagrange multipliers, and reduce the
canonical bracket to the variables $Q^r$ and $P_r$. The dynamics for
this reduced set of variables will remain unchanged. So we are led to a
dynamics described by
\beq
H_R(Q^r,P_r) + Q^{a_2} f_{a_2} (Q^r,P_r),
\eeq
with $Q^{a_2}$ now playing the role of Lagrange multipliers. At this
point we have arrived at the same result obtained after performing the
first step in F-J method, (\ref{LR}).

To summarize, one step in F-J method corresponds to one step and a half
in Dirac's, plus the adoption of Dirac brackets after each step, plus
the trading of the remaining Lagrange multipliers (here $\eta^{a_2}$) by
a set of variables (here $Q^{a_2}$) that in fact are its primitives,
according to (\ref{primit}). {\bf This proves the full equivalence
of both methods}.

\section{Example}

In particular examples it may be difficult to compare the results
obtained by the two reduction procedures, that of Dirac and F-J
approaches.  It may happen that the reduced Lagrangian obtained from
Dirac reduction coincide exactly with the reduced Lagrangian obtained
from F-J reduction procedure --as in the case when we have only primary
second class constraints--.  Other possible instance is that the
resulting Lagrangians can be related by a canonical transformation --as
in the case when we choose some gauge fixing constraints in Dirac's
reduction in a different way that the natural choice implicit in the F-J
analysis, namely by setting to zero all the gauge degrees of freedom--.
The situation may be still more involved and the two reduced
Lagrangians may differ by a non-canonical transformation --as in the
case when, at some stage of F-J procedure, the set of constraints
$(\ref{zeta2})$ is empty--.  This situation can arise because enforcing
the constraints $(\ref{zeta1})$ through F-J method, does not alter the
symplectic structure of the Lagrangian $(\ref{L'})$, while the Dirac
formalism with second class constraints gives Dirac brackets that, in
general, are not diagonal.  In any case we have always the possibility
to resort to the equivalence proof given in section 2 to analyze the
explicit form of the transformation that connects the two reduced
Lagrangians.  The situation can be described by the following diagram
\[
\begin{array}{ccc}
L(q,p)&
\stackrel{DIRAC-REDUCTION}{\longrightarrow}&
L_{R}(q_r, p_r)\\
\vcenter{%
\llap{$\scriptstyle{{\rm CANONICAL}}$}}\Bigg\downarrow  & &
\Bigg\downarrow\vcenter{%
\rlap{$\scriptstyle{{\rm  NON-CANONICAL}}$}}\\
{\bar L}(Q,P)&
\stackrel{F-J-REDUCTION}{\longrightarrow}&
{\bar L}_{R}( Q_r, P_r)
\end{array}
\]
where the non-canonical transformation can be calculated by the
restriction to the constraint surface of the canonical transformation
that relates the two Lagrangians, as we show in the following example.
Notice that this non-canonical transformation arises because in the
F-J reduction process its necessary to perform a Darboux transformation
at each level of the algorithm.

The aim of this section is to illustrate some of the ideas
of the previous paragraphs by means of an example that exhibits this
type of behavior.
To this end let us take a dynamical  system represented by the
following first order Lagrangian \cite{Sun}.
\beq
L=p_1{\dot q}_1+p_2{\dot q}_2-H_c(q,p)-\lambda\phi_1
\label{eL}
\eeq
where
\beq
H_c=\frac12(p_1-q_2)^2-\frac\beta2(q_1-q_2)^2,
\quad\phi=p_2-(1-\alpha)q_1,
\eeq
with $\alpha \neq \beta$. The stabilization of $\phi$ leads to a
secondary constraint
\beq
\phi_2=\alpha(p_1-q_2)-\beta(q_1-q_2)=0
\label{c1}
\eeq
which upon stabilization gives
\beq
\alpha\beta(q_1-q_2)-\beta(p_1-q_2)-\gamma\lambda=0
\label{c2}
\eeq
where $\gamma\equiv\alpha^2-\beta$. The constraints are second class if
$\gamma\neq 0$ and
first class when $\gamma=0$.
In the first case $(\gamma\neq 0)$ the
implementation of the constraints leads to the reduced Lagrangian
\beq
L_{R}=\frac{\gamma}{\alpha-\beta}
{\dot q}_1p_1+\frac12
\frac{\beta\gamma}{(\alpha-\beta)^2}(p_1-q_1)^2.
\label{LRD}
\eeq
Other equivalent reductions can be implemented. Here we choose to
eliminate $q_2,p_2$ from the constraints (\ref{c1}) and (\ref{c2}).
The Dirac brackets are
\begin{eqnarray}
\{q_1,q_2\}_D=\frac\alpha\gamma , \quad
\{q_1,p_1\}_D= \frac{\alpha-\beta}{\gamma} ,
\quad\{q_2,p_2\}_D=\frac{\alpha(1-\alpha)}{\gamma} , \\
\{p_1,p_2\}_D=\frac{(1-\alpha)(\beta-\alpha)}{\gamma} ,
\quad \{q_1,p_2\}_D= 0 ,
\quad\{q_2,p_1\}_D=-\frac\beta\gamma.
\end{eqnarray}
Note that the Dirac brackets in reduced phase space can be read out
directly from the reduced Lagrangian $(\ref{LRD})$.

In the second case $(\gamma=0)$ the standard reduction procedure can be
implemented through a gauge fixing.  The natural choice is $q_2=0$ and
$q_1=0$ and the reduced Lagrangian is zero, as expected.  A general
gauge fixing of the form $q_1=\chi_1(q,p), q_2=\chi_2(q,p)$, where
$\chi_1$ does not depends on $q_1$ and $\chi_2$ not depends on $q_2$,
which we suppose that fixes the gauge without ambiguities gives, in
general, a reduced Lagrangian that is a total time derivative of some
function.  This fact means that the two gauge fixing procedures gives
first order reduced Lagrangians that differ by a canonical
transformation.

Let us perform the F-J analysis for this system. In the $(\gamma\neq 0)$ case
the first step (implementing $\phi_1=0$) of F-J method gives the reduced
Lagrangian
\beq
L={\dot q}_1(p_1-(1-\alpha)q_2)-H_c(q,p),
\eeq
up to a total time derivative. We can diagonalize the symplectic structure by
means of the Darboux transformation
\beq
P_1=p_1-(1-\alpha)q_2\quad Q_1=q_1\quad Q_2=q_2
\label{DT}
\eeq
which leads to
\beq
L'={\dot Q}_1P_1-\frac12(P_1-\alpha Q_2)^2 +\frac{\beta}{2}(Q_1-Q_2)^2
\label{LQP}
\eeq
We note that $Q_2$ plays the role
of a $Z$-type variable. From the condition $(\ref{H'})$ it follows that
\beq
Q_2=\frac{\alpha P_1-\beta Q_1}{\gamma}.
\eeq
We deduce from the general analysis that this
constraint is going to play the role of a second class constraint.
The elimination of $Q_2$ from the Lagrangian $(\ref{LQP})$ gives
\beq
L_{R}={\dot Q}_1P_1+\frac{\beta}{2\gamma}(P_1-\alpha Q_1)^2.
\label{eLRFJ}
\eeq
That this Lagrangian does not coincide with
the corresponding Lagrangian $(\ref{LRD})$ obtained via Dirac reduction
procedure may be surprising at first sight. In fact these two
Lagrangians are
related via a non-canonical transformation that correspond to the restriction
of the canonical transformation
\beq
(q_1,q_2,p_1,p_2)= (Q_1,Q_2,P_1+(1-\alpha)Q_2, P_2+(1-\alpha)Q_1),
\eeq
allowing for  a canonical representation of
the constraint surface, to the surface
$P_2=0,Q_2-\frac{\alpha P_1-\beta Q_1}{\gamma}=0$. Note that this
transformation is a particular extension off the constraint surface of the
Darboux transformation $(\ref{DT})$.
The non-canonical transformation that relates this two Lagrangians is thus
\beq
(q_1,p_1)=(Q_1,P_1+\frac{(1-\alpha)(\alpha P_1-\beta Q_1)}{\gamma}).
\eeq
By applying this transformation to the Dirac Lagrangian $(\ref{LRD})$
we recover precisely the result obtained via F-J method $(\ref{eLRFJ})$.

The case of first class constraints, when $\gamma=0$, can be
calculated in a similar way. The first reduction gives
\beq
L={\dot Q}_1P_1-\frac12 P_1^2+\frac{\alpha^2}{2} Q_1^2+Q_2(\alpha
P_1-\alpha^2Q_1),
\label{FCL}
\eeq
after a proper diagonalization via the Darboux
transformation $(\ref{DT})$.
Note that the Lagrangian $(\ref{FCL})$ has already the form $(\ref{LR})$
where
$Q_2$ play, form now, the role of a Lagrange multiplier. The secondary
constraint,
\beq
P_1-\alpha Q_1 = 0,
\label{SC}
\eeq
does not allow to obtain $Q_2$ as function of the rest of
coordinates.
As we noted, this is the condition that enable us to classify this
secondary constraint as a first class constraint. The implementation
of $(\ref{SC})$ on the Lagrangian $(\ref{FCL})$ gives a
Lagrangian which is a total
time derivative. We then conclude that the reduced Lagrangian obtained
via Dirac reduction and that obtained via F-J reduction are totally
equivalent.
They differ at most by a canonical transformation.

\section{Conclusions}

We have proved the full equivalence between the Dirac approach and the F-J
analysis for constrained systems.
While getting this equivalence we have obtained new insights into the
F-J method. Let us quote some.

{\it First}.  We have identified the constraints involving the $Z_{a_1}$
(or $Q^{a_1}$, see $(\ref{zeta1})$ and $(\ref{split})$)  variables as the subset
of secondary second class
constraints that bring a subset of the primary first class constraints
into second class.  Here is where the old Dirac's classification of
constraints into first and second class is still present in F-J method.

{\it Second}. We have said that it may be that some --or all-- of the
relations $f_{a_2}(P_r,Q^r) = 0$ (which are the secondary constraints
that do not alter the first or second class character of the primary
constraints) are empty.  From the point of view of Dirac's method, this
happens when the number of --independent-- secondary constraints is less
than the number of initial primary first class constraints.  Also,
from the point of
view of the construction of the generators of the gauge transformations
(which are made up with first class constraints in a chain involving an
arbitrary function and its time derivatives), this means that there are
as many chains made up with a single constraint as the difference
between the two numbers just mentioned.

{\it Third}.  Notice that if we start the second step in F-J method
$(\ref{LR})$,
by plugging the constraints $f_{a_2} (Q^r,P_r) = 0$ into $L_R$, all the
information on the variables $Q^{a_2}$ disappears.  So in F-J method we
will never encounter any equation for these variables.  Instead, in our version
of
Dirac's approach, since these variables will play the role of Lagrange
multipliers, either they are finally determined as functions of the rest
of variables $P_r,Q^r$,
or remain arbitrary (and hence, gauge)
functions.  At this point we do not know the particular fate --among
these two possibilities-- of each of these variables but, no matter
which one will it be, we already know that these variables do not carry
any relevant dynamical information and they will never account for
physical degrees of freedom.  Here lies the nice efficiency of F-J
method we have mentioned at the introduction: getting rid of these
variables as soon as possible, instead of carrying them on board until
the end, as it is usually done with Dirac's method.

{\it Fourth}.  From a geometrical point of view the procedure leading
to the new phase space variables $P_i,Q^i$, beginning with the canonical
representation of the constraint surface (\ref{M}) and ending with the
canonical transformation (\ref{cr}), is equivalent to extend off the
constraint surface the Darboux transformation (\ref{darboux}) needed in
the F-J approach, as we have seen.  This can be visualized as follows.
We can perform a transformation of coordinates from the original phase
space variables $p_i,q^i$ to new ones $x^s, \phi_\mu$ where $q^i(x^s),
p_i(x^s)$ are the parametric equations of the constraint surface and
$\phi_\mu=0$ are the constraints.  By the application of this
transformation to the original canonical symplectic form we obtain a new
symplectic form that in general is not diagonal.  Now by means of a
Darboux transformation we can diagonalize this symplectic form and
obtain canonical coordinates and some variables of type $Z$.  The new
symplectic form is diagonal in a reduced phase space.  Extending off the
constraint surface this Darboux transformation results in the canonical
transformation $(\ref{cr})$ used in the text.

In some cases when the matrix $M_\mu^\nu$ in equation (\ref{M}) is easy
to find, the procedure presented here can be used as an alternative way
to construct the Darboux transformation which is needed in the F-J
approach.  From this point of view the Dirac method plus the canonical
transformation $(\ref{cr})$ that diagonalizes the Dirac brackets in a
stage by stage procedure reduces to the F-J approach.

{\it Fifth.} As we have already pointed out, the F-J method is an
efficient way to achieve the reduction and get the physical variables in
the reduced phase-space formulation.  The method is technically
different but the result as compared with the standard reduced
formulation is the {\it same}: If we introduce a gauge-fixing to
eliminate the gauge degrees of freedom in Dirac's method, we will obtain
the same reduced formulation.  So, as a starting point to quantization,
the F-J approach presents for a general case all the problems
inherent to the classical
reduction of variables (posible loss of covariance, locality, or
anomalies, etc.) prior to quantization.

\section*{Acknowledgements}

We would like to thank J. Gomis for reading and comments on the
manuscript.  J.M.P. acknowledges support by the CICIT (contract numbers
AEN95-0590 and GRQ93-1047) and by a Human Capital and Mobility Grant
(ERB4050PL930544).  J.A.G. is partially supported by CONACyT graduate
fellowship \# 86226 and also thanks the Departament d'Estructura i
Constituents de la Mat\`eria at the Universitat de Barcelona for its
hospitality.

\end{document}